\documentclass[aps,amsmath,amssymb,twocolumn,superscriptaddress,nofootinbib]{revtex4-1}

\usepackage{graphicx}
\usepackage{dcolumn}
\usepackage{bm}
\usepackage{color}
\usepackage[colorlinks=true,pdfstartview=FitV,linkcolor=blue,citecolor=blue,urlcolor=blue]{hyperref}
\usepackage[mathlines]{lineno}
\usepackage[dvipsnames]{xcolor}
\usepackage{amsmath}

\everymath{\displaystyle}
\begin{document}

\newcommand{\up}[1]{$^{#1}$}
\newcommand{\down}[1]{$_{#1}$}
\newcommand{\powero}[1]{\mbox{10$^{#1}$}}
\newcommand{\powert}[2]{\mbox{#2$\times$10$^{#1}$}}

\newcommand{\evm}{\mbox{\rm{eV\,$c^{-2}$}}}
\newcommand{\mevm}{\mbox{\rm{MeV\,$c^{-2}$}}}
\newcommand{\gevm}{\mbox{\rm{GeV\,$c^{-2}$}}}
\newcommand{\pgd}{\mbox{g$^{-1}$\,d$^{-1}$}}
\newcommand{\um}{\mbox{$\mu$m}}
\newcommand{\spix}{\mbox{$\sigma_{\rm pix}$}}
\newcommand{\pav}{\mbox{$\langle p \rangle$}}

\newcommand{\sige}{\mbox{$\bar{\sigma_e}$}}
\newcommand{\mass}{\mbox{$m_\chi$}}
\newcommand{\crystal}{\mbox{$f_\textnormal{c}(q,E_e)$}}
\newcommand{\electron}{\mbox{$\rm{e^-}$}}

\title{Constraints on Light Dark Matter Particles Interacting with Electrons from DAMIC at SNOLAB}

\author{A.~Aguilar-Arevalo} 
\affiliation{Universidad Nacional Aut{\'o}noma de M{\'e}xico, Mexico City, Mexico} 

\author{D.~Amidei}
\affiliation{Department of Physics, University of Michigan, Ann Arbor, MI, United States}  

\author{D.~Baxter}
\affiliation{Kavli Institute for Cosmological Physics and The Enrico Fermi Institute, The University of Chicago, Chicago, IL, United States}

\author{G.~Cancelo}
\affiliation{Fermi National Accelerator Laboratory, Batavia, IL, United States}

\author{B.A.~Cervantes Vergara}
\affiliation{Universidad Nacional Aut{\'o}noma de M{\'e}xico, Mexico City, Mexico} 

\author{A.E.~Chavarria}
\affiliation{Center for Experimental Nuclear Physics and Astrophysics, University of Washington, Seattle, WA, United States}

\author{E.~Darragh-Ford}
\affiliation{Kavli Institute for Cosmological Physics and The Enrico Fermi Institute, The University of Chicago, Chicago, IL, United States}

\author{J.R.T.~de~Mello~Neto}
\affiliation{Universidade Federal do Rio de Janeiro, Instituto de  F\'{\i}sica, Rio de Janeiro, Brazil}

\author{ J.C.~D'Olivo}
\affiliation{Universidad Nacional Aut{\'o}noma de M{\'e}xico, Mexico City, Mexico} 

\author{J.~Estrada}
\affiliation{Fermi National Accelerator Laboratory, Batavia, IL, United States}

\author{R.~Ga\"ior}
\affiliation{Laboratoire de Physique Nucl\'eaire et des Hautes \'Energies (LPNHE), Sorbonne Universit\'e, Universit\'e de Paris, CNRS-IN2P3, Paris  France}

\author{Y.~Guardincerri}
\thanks{Deceased January 2017}
\affiliation{Fermi National Accelerator Laboratory, Batavia, IL, United States}

\author{ T.W.~Hossbach}
\affiliation{Pacific Northwest National Laboratory (PNNL), Richland, WA, United States} 

\author{B.~Kilminster}
\affiliation{Universit{\"a}t Z{\"u}rich Physik Institut, Zurich, Switzerland }

\author{I.~Lawson}
\affiliation{SNOLAB, Lively, ON, Canada }

\author{S.J.~Lee}
\affiliation{Universit{\"a}t Z{\"u}rich Physik Institut, Zurich, Switzerland }

\author{A.~Letessier-Selvon}
\affiliation{Laboratoire de Physique Nucl\'eaire et des Hautes \'Energies (LPNHE), Sorbonne Universit\'e, Universit\'e de Paris, CNRS-IN2P3, Paris  France}

\author{A.~Matalon}
\affiliation{Kavli Institute for Cosmological Physics and The Enrico Fermi Institute, The University of Chicago, Chicago, IL, United States}
\affiliation{Laboratoire de Physique Nucl\'eaire et des Hautes \'Energies (LPNHE), Sorbonne Universit\'e, Universit\'e de Paris, CNRS-IN2P3, Paris  France}

\author{V.B.B.~Mello}
\affiliation{Universidade Federal do Rio de Janeiro, Instituto de  F\'{\i}sica, Rio de Janeiro, Brazil}

\author{P.~Mitra}
\affiliation{Center for Experimental Nuclear Physics and Astrophysics, University of Washington, Seattle, WA, United States}

\author{J.~Molina}
\affiliation{Facultad de Ingenier\'{\i}a, Universidad Nacional de Asunci\'on, Asuncion, Paraguay}

\author{S.~Paul}
\affiliation{Kavli Institute for Cosmological Physics and The Enrico Fermi Institute, The University of Chicago, Chicago, IL, United States}

\author{A.~Piers}
\affiliation{Center for Experimental Nuclear Physics and Astrophysics, University of Washington, Seattle, WA, United States}

\author{P.~Privitera}
\affiliation{Kavli Institute for Cosmological Physics and The Enrico Fermi Institute, The University of Chicago, Chicago, IL, United States}
\affiliation{Laboratoire de Physique Nucl\'eaire et des Hautes \'Energies (LPNHE), Sorbonne Universit\'e, Universit\'e de Paris, CNRS-IN2P3, Paris  France}

\author{K.~Ramanathan}
\affiliation{Kavli Institute for Cosmological Physics and The Enrico Fermi Institute, The University of Chicago, Chicago, IL, United States}

\author{J.~Da~Rocha}
\affiliation{Laboratoire de Physique Nucl\'eaire et des Hautes \'Energies (LPNHE), Sorbonne Universit\'e, Universit\'e de Paris, CNRS-IN2P3, Paris  France}

\author{ Y.~Sarkis}
\affiliation{Universidad Nacional Aut{\'o}noma de M{\'e}xico, Mexico City, Mexico} 

\author{M.~Settimo}
\affiliation{SUBATECH, CNRS-IN2P3, IMT Atlantique, Universit\'e de Nantes, Nantes, France}

\author{R.~Smida}
\affiliation{Kavli Institute for Cosmological Physics and The Enrico Fermi Institute, The University of Chicago, Chicago, IL, United States}

\author{R.~Thomas}
\affiliation{Kavli Institute for Cosmological Physics and The Enrico Fermi Institute, The University of Chicago, Chicago, IL, United States}

\author{J.~Tiffenberg}
\affiliation{Fermi National Accelerator Laboratory, Batavia, IL, United States}

\author{D.~Torres Machado}
\affiliation{Universidade Federal do Rio de Janeiro, Instituto de  F\'{\i}sica, Rio de Janeiro, Brazil}

\author{R. Vilar}
\affiliation{Instituto de F\'isica de Cantabria (IFCA), CSIC \-- Universidad de Cantabria, Santander, Spain}

\author{A.L.~Virto}
\affiliation{Instituto de F\'isica de Cantabria (IFCA), CSIC \-- Universidad de Cantabria, Santander, Spain}

\collaboration{DAMIC Collaboration}
\noaffiliation

\date{\today}

\begin{abstract}

We report direct-detection constraints on light dark matter particles interacting with electrons. The results are based on a method that exploits the extremely low levels of leakage current of the DAMIC detector at SNOLAB of \powert{-22}{2\textendash6}~A\,cm$^{-2}$. We evaluate the charge distribution of pixels that collect $<10~\rm{e^-}$ for contributions beyond the leakage current that may be attributed to dark matter interactions. Constraints are placed on so-far unexplored parameter space for dark matter masses between 0.6 and 100 \mevm. We also present new constraints on hidden-photon dark matter with masses in the range $1.2$\textendash$30$~\evm. 
\end{abstract}


\maketitle
There is overwhelming astrophysical and cosmological evidence for Dark Matter (DM) as a major constituent of the universe. Still, its nature remains elusive. The compelling Weakly Interacting Massive Particle (WIMP) dark matter hypothesis \cite{kolb:1990vq} \textemdash\ implying DM is made of hitherto unknown particles with mass in the GeV\textendash TeV scale \textemdash\ has been intensely scrutinized during the last two decades by detectors up to the tonne-scale looking for nuclear recoils induced by coherent scattering of WIMPs. Despite the impressive improvements in sensitivity, notably by noble liquid experiments \cite{{aprile:2018, *akerib:2017, *agnes:2018}}, WIMPs have so far escaped detection. Other viable candidates include DM particles from a hidden-sector \cite{battaglieri2017us}, which couple weakly with ordinary matter through, for example, mixing of a hidden-photon with an ordinary photon \cite{{okun:1982, *holdom:1985}}. A phenomenological consequence is that hidden-sector DM particles also interact with electrons, with sufficiently large energy transfers to be detectable down to DM masses of $\approx$~MeV \cite{essig:2012}. Also, eV-mass hidden-photon DM particles can be probed through absorption by electrons in detection targets \cite{hochberg:2017}. 

The DAMIC (Dark Matter in CCDs) experiment \cite{Aguilar-Arevalo:2016} is well-suited for a sensitive search of this class of DM candidates. DAMIC detects ionization events induced in the bulk silicon of thick, fully depleted Charge-Coupled Devices (CCDs). By exploiting the charge resolution of the CCDs ($\approx 2$~\electron) and their extremely low leakage current ($\approx 4$~\electron\,$\rm{mm^{-2}}\,\rm{d^{-1}}$), DAMIC has already placed constraints on hidden-photon DM with masses in the range $1.2$\textendash$30$~\evm\ \cite{aguilar:2017} with data collected in a preliminary science run. In this Letter we apply a similar approach to explore DM-\electron\ interactions with high-quality data from the DAMIC science run at the SNOLAB underground laboratory. We also present improved limits on hidden-photon DM particles.

To model DM-\electron\ interactions we follow Ref. \cite{essig:2016} where the bound nature of the electrons and crystalline band structure of the target are properly taken into account. The differential event rate in the detector for a DM mass \mass, with transferred energy $E_e$, and momentum $q$ is parametrized as
\begin{equation} \label{eq:rate}
\frac{dR}{dE_e} \propto  \bar{\sigma_e} \int \frac{dq}{q^2}~\eta(m_\chi,q,E_e) |F_{DM}(q)|^2 |f_\textnormal{c}(q,E_e)|^2\,,
\end{equation} 
where \sige\ is a reference cross section for free electron scattering, $\eta$ includes properties of the incident flux of galactic DM particles, $F_{DM}$ is the dark matter form factor, and the crystal form factor \crystal\ quantifies the atomic transitions of bound-state electrons.

The DM form factor expresses the momentum-transfer dependence of the interaction, generically introduced as $F_{DM}=(\alpha m_e/q)^n$ \{$n=0,1,2$\}. The $n=0$ case corresponds to point-like interactions with heavy mediators or a magnetic dipole coupling, the $n=1$ case to an electric dipole coupling, and $n=2$ to massless or ultra-light mediators. The crystal form factor encodes target material properties and is calculated numerically from a density functional theory (DFT) approach, with results taken from Ref. \cite{essig:2016} for silicon.

The DAMIC detector has taken data at SNOLAB since 2017 with seven CCDs (4k $\times$ 4k-pixel, 15 $\times$ 15~\um$^2$ pixel size, 675~\um~thick for 6.0~g mass each). The devices are fully depleted and a drift field is maintained across the CCD thickness by the application of 70~V to a thin backside contact. The CCDs are operated at $\approx140$~K (stable to within $0.5$~K) inside a copper vacuum vessel kept at $\sim$\powero{-6}~mbar. The CCD tower is shielded on all sides by at least 18~cm of lead, with the innermost 5~cm of ancient origin, and 42~cm of polyethylene to stop background radiation from environmental $\gamma$ rays and neutrons, respectively. Each CCD is read out serially by three-phase clocking, which first moves the charge in rows of pixels vertically ($y$-direction) into the serial register. Then, single pixels are shifted horizontally ($x$-direction) into the readout node, a charge to voltage amplifier located at a corner of the device. A second readout node at the opposite end of the serial register is also read out synchronously, providing a correlated noise-only measurement. An analog-to-digital converter (ADC) measures the readout node voltage, giving a pixel value $p$ in analog-to-digital converter units (ADU) linearly proportional to the number of charges in the pixel. The CCDs are individually calibrated in-situ by a red LED, with conversion constants $\Omega \approx 14.5$~ADU/\electron. The standard mode of data taking consists of 30~ks ($\approx$~8.3 hours) long exposures followed by readout. ``Blank" images with a much shorter 30~s exposure are also taken immediately after each long exposure as a systematic check of the device operation. Details of device architecture, DAMIC infrastructure, calibration, and image processing are given in Refs. \cite{Aguilar-Arevalo:2016, aguilar:2017}.

The  search reported here was performed on a special data set consisting of 38 exposures, each 100~ks ($\approx$~1.16~days) long, collected in late 2017. This longer exposure time allows for a more precise determination of the leakage current. The data were acquired with 1\,$\times$\,100 binning, a readout mode where the charges of 100 consecutive pixels in a column are summed into the serial register before readout. The binned pixel size is thus 15\,$\times $\,1500~\um$^2$. Since readout noise is introduced each time the charge is measured, a better signal-to-noise ratio in the measurement of the charge collected over multiple pixels is achieved by binning. Hereafter, the term pixel will refer to a binned pixel. Each image contains 4272\,$\times$\,193 pixels, with a subset of 4116\,$\times$\,42 pixels corresponding to the active area of the CCD. The extra pixels, referred to as the $x$ and $y$ overscans, do not contain any charge since they are the result of clocking the CCD past the active region.
\begin{figure}[t!]
	\centering
	\includegraphics[width=0.48\textwidth]{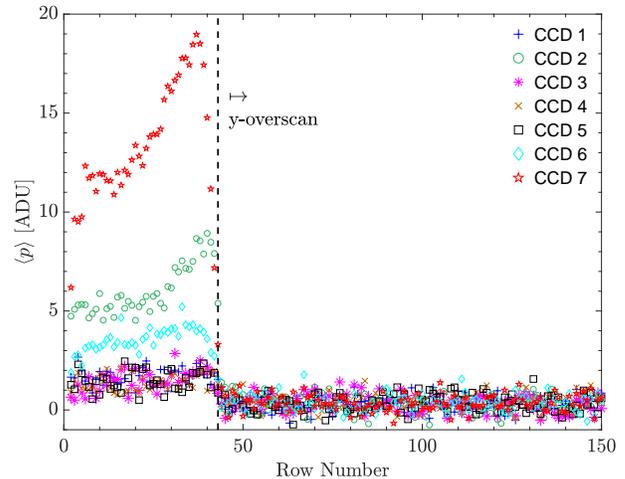}
	\caption{Mean pixel ADU values, after the processing described in the text, as a function of row in the CCD. The first 42 rows correspond to the active region of the CCD, while rows $\ge 43$ correspond to the $y$ overscan. The offset observed in rows $\le42$ is due to charge accumulated in the pixels.}
	\label{fig:ext212}
\end{figure}

Image processing begins with subtraction of the constant offset (``pedestal") present in each pixel introduced by the electronics chain. The pedestal is estimated on a per-row basis as the mean value of pixels in the $x$-overscan. To exclude an instrumental increase in transient noise at the boundaries of the CCDs, the analysis is restricted to 2500 columns in a central portion of the image. To remove correlated readout noise, we subtract from every pixel an appropriate linear combination of corresponding pixel values in the noise images acquired with the aforementioned second readout node. The subtraction coefficients are calculated to minimize the variance of the pixel noise. The resulting image noise is found to be $ \sigma_{pix} \approx 1.6$ \electron\ as reported in Table \ref{tab:fits}. 

Physical defects in the silicon lattice structure of the CCDs often result in localized regions of high dark current, generating hot pixels and columns recurring over multiple images. A mask obtained from a statistical analysis of 864 images of the full-science data set (see details of the methodology in Ref. \cite{aguilar:2017}) is applied, resulting in the removal of $\approx$0.25\% of pixels.

Clusters of pixels with signal larger than 8\,$\sigma_{pix}$, arising from ionization events by particles \cite{Aguilar-Arevalo:2016} that deposit more than $50$~eV, are also excluded as to limit the analysis to leakage current and signals from light dark matter. To mitigate the effect of charge trailing along rows from charge transfer inefficiency in the serial register, 200 pixels to the left of every cluster are masked along with 4 pixels to the right. Each pixel above and below these clusters is also masked to account for charge splitting across rows due to diffusion. This procedure removes $\approx$ 2.0\% of pixels.

\begin{figure}[!h]
	\centering
	\includegraphics[width=0.48\textwidth]{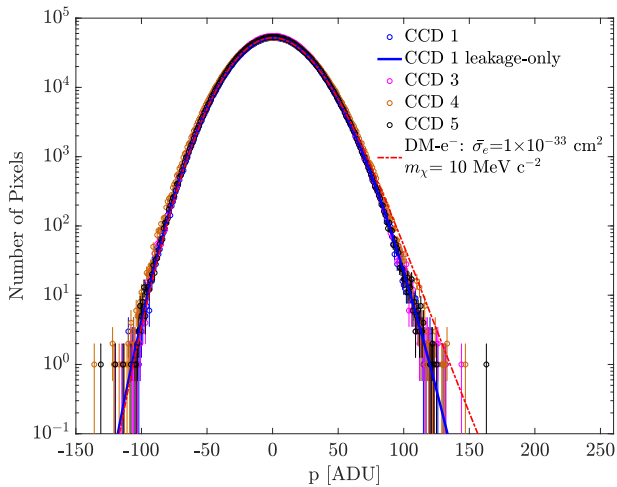}
	\caption{Distribution of pixel values (with aforementioned conversion constants $\Omega \approx 14.5$~ADU/\electron) for the four CCDs selected for this analysis. An example of best fit result for the leakage-only model (no DM-\electron) is given for CCD n. 1 (blue line); the dashed red line is the expectation for a DM-\electron\ model with \sige\ = $1 \times 10^{-33}$ cm$^2$, \mass\ = 10 \mevm and $F_{DM}=1$. 
	}
	\label{fig:exthists}
\end{figure}

After applying these image processing and pixel selection procedures, we calculate the mean value of pixels $\langle p \rangle$ in each row over the 38 images of the data set (Fig. \ref{fig:ext212}). Rows 43 and higher correspond to the $y$ overscan, where $\langle p \rangle$ is consistent with zero. CCD data are contained in the first 42 rows of the image, where an offset is clearly present due to charge collected by the pixels. CCD numbers 2, 6 and 7 present a significantly higher leakage current that is non-uniform across the rows. This is likely due to external sources \textemdash\ e.g. optical or IR photons in the vessel \textemdash\ and inconsistent with DM which would produce charge uniformly distributed throughout the pixel array. Thus these CCDs are not considered any further in this analysis. For the four remaining CCDs, the analysis is restricted to rows 1-36 where $\langle p \rangle$ is found to be constant within uncertainty. The final selected region includes $\approx$ \powert{6}{3.2}~pixels for each of the four CCDs, with their corresponding pixel value distributions shown in Fig. \ref{fig:exthists}. The total equivalent exposure of the search is 200 g\,d.

The distribution of pixel values in a CCD is shown in Fig.~\ref{fig:exthists} and is modeled by the function $\Pi(p)$, which comes from the convolution of the pixel charge with the pixel readout noise. We take the pixel charge to be the sum of a Poisson-distributed leakage current $\lambda$ accumulated during the exposure and a DM signal $S$ derived from Eq.~\ref{eq:rate}, where $S \equiv S(j \mid \bar{\sigma_e}, m_\chi)$ specifies the probability to produce $j$ charges in a pixel from specific DM interactions. The readout noise is parametrized from the pixel value distribution of blanks and overscans, and found to be well-described by the convolution of a Poisson with average $\lambda_{d}$ and a Gaussian of standard deviation $\sigma_{\rm pix}$. This parametrization reflects the presence of non-Gaussian features in the noise distribution. The pixel value distribution for a given CCD is then derived as:
\begin{widetext}
\begin{align} \label{eq:pixdist}
\begin{split}
\Pi(p) &= N \sum_{n_c=0}^{\infty} \sum_{n_l=0}^{\infty}\bigg( \bigg[ \sum_{j=0}^{n_c} S(j \mid \bar{\sigma_e}, m_\chi) \textnormal{Pois}(n_c-j \mid \lambda) \bigg] \textnormal{Pois}(n_l \mid \lambda_{d}) \textnormal{Gaus}(p \mid \Omega \big[(n_c + n_l) + \mu_0 \big], \Omega \sigma_{\rm pix})\, \bigg) \\
&= N \sum_{n_{\rm tot}=0}^{\infty}  \bigg( \bigg[ \sum_{j=0}^{n_{\rm tot}} S(j \mid \bar{\sigma_e}, m_\chi) \textnormal{Pois}(n_{\rm tot}-j \mid \lambda_{\rm tot}) \bigg]  \textnormal{Gaus}(p \mid \Omega \big[n_{\rm tot} + \mu_0 \big], \Omega \sigma_{\rm pix})\, \bigg)\,,
\end{split}
\end{align}
\begin{equation} \label{eq:pixdist1}
\rm{with}~~\textit{n}_{\rm{tot}} = \textit{n$_c$}+\textit{n$_l$} ~;~~ \lambda_{\rm tot} = \lambda_d+\lambda\,,
\end{equation}
\end{widetext}
where $N$ is the number of pixels in the dataset, $n_{c}$ is the number of charges in a pixel from the DM signal and leakage current, $n_{l}$ is the number of charges in a pixel from readout shot noise, $\Omega$ is the \electron to ADU calibration constant, and $\mu_0$ is an offset accounting for pedestal subtraction.
The noise parameters $\sigma_{pix}$, $\lambda_d$ and $\mu_0$ reported in Table \ref{tab:fits} are determined from a fit of the blanks and $y$-overscans. We then perform a maximum likelihood fit of the data to the leakage-only model (i.e. no contribution from DM-\electron\ interactions, corresponding to  $S(0)=1$ and $S(j\ge1)=0$) with $\sigma_{pix}$ and $\mu_0$ constrained with Gaussian penalty terms. The leakage current parameter $\lambda$
derived from the leakage-only best-fit value of $\lambda_{\rm tot}$ is reported in Table \ref{tab:fits}; $\sigma_{pix}$ and $\mu_0$ from the constrained fit were found to be consistent with the blank and $y$-overscan values. Notice that $\lambda$ represents an upper limit to the leakage current, with $\lambda=1.0$~e$^-$mm\,$^{-2}$\,d$^{-1}$ ($\approx$ \powert{-22}{2}~A\,cm$^{-2}$) for CCD 4, the lowest ever measured in a silicon device.

\begin{table}[h!]
	\caption{\label{tab:fits} Relevant parameters used in modeling the pixel value distribution, with statistical uncertainty in parentheses. The first three columns correspond to the fit of blanks and overscans, while the last column to the leakage-only fit to data. Where appropriate, units were converted from e$^-$\,pix$^{-1}$\,img$^{-1}$, as for Eq. \ref{eq:pixdist}, to e$^-$\,mm$^{-2}$\,d$^{-1}$.}
	\begin{ruledtabular}
		\begin{tabular}{ccccc}
			CCD n.	& $\sigma_{pix}$			& $\lambda_{d}$	& $\mu_{0}$	&	$\lambda = \lambda_{tot}-\lambda_d$							\\
				& [e$^-$] 				&[e$^-$\,mm$^{-2}$\,img$^{-1}$]		& 	[e$^-$]	 & [e$^-$\,mm$^{-2}$\,d$^{-1}$]	\\				
				\hline
		1  &  1.628(1)  &  ~8.2(2)   &  -0.185(3)  & 2.8(2)  	\\
		3  &  1.572(1)  &  ~7.8(2)  &  -0.160(4)  &  1.7(2)  	\\
		4  &  1.594(1)  & 10.0(2)  &  -0.219(4)  & 1.0(2)  	\\
		5  &  1.621(1)  &  ~8.5(2)  &  -0.183(4)  & 2.0(2)  	\\
		\end{tabular}

	\end{ruledtabular}
\end{table}

The DM signal is computed using Eq. \ref{eq:rate}. We obtain the distribution of \crystal\ from the binned output of the \textit{QEDark} \cite{essig:2016, priv_comm_sensei} module written for the \textit{QuantumEspresso} \cite{quantumespresso} DFT code. To compute $\eta$ we assume halo parameters of dark matter density $\rho_{DM} = 0.3$ \gevm\ cm$^{-3}$, an isothermal Maxwellian velocity distribution with escape velocity v$_{\rm{esc}}$ = 544~km\,s$^{-1}$ and mean v$_0$ = 220~km\,s$^{-1}$, and periodic Earth motion with mean velocity v$_E$ = 232~km\,s$^{-1}$ \cite{lewin:1996}. The resulting ionization rate $dR/dE_e$ is then discretized into $dR/dn_e$, where $n_e$ is the number of ionization charges. For this purpose we use Monte-Carlo-derived probabilities $P(n_e \vert E_e)$ to produce electron-hole (e-h) pairs, informed from studies in Ref. \cite{{alig:1980, *alig1983scattering}}, with the assumption that the initial energy deposit is split randomly between the e-h pair. Measurements of direct charge injection \cite{chang:1985,scholze1998} validate the quantum yield of these prescriptions for deposits $<5$~eV; these prescriptions also match the Fano factor \cite{fano:1947} measured with similar CCDs in Ref. \cite{compton:2017}. The ionization rate is then obtained from $dR/dn_e = \int dE_e P(n_e \vert E_e)(dR/dE_e)$. Lastly, the effect of charge diffusion in the CCDs is included. In fact, a point-like charge deposit in the silicon bulk of the CCD may split over several pixels due to diffusion of the ionized charge as it drifts towards the pixel array. To derive the effective signal distribution, point-like charge deposits uniformly distributed across the depth $z$ of the CCD are simulated according to $dR/dn_e$. The charges are then distributed in the $x$-$y$ pixel array following the spatial variance $\sigma_{xy}^2(z)$ from a diffusion model derived from data \cite{Aguilar-Arevalo:2016}, and a distribution of charges collected by a pixel is obtained. The procedure is repeated 1000 times to obtain the numerical distribution for the DM signal $S(j \mid \bar{\sigma_e}, m_{\chi})$. Examples of the DM model and leakage-only expectations are shown in Fig. \ref{fig:exthists}.

To constrain the DM signal, we implement a likelihood analysis in (\sige\,\mass) space. For a fixed \mass\ and for every CCD $i$ we minimize the negative log-likelihood $\mathcal{LL}_i$ of $\Pi(p)$, leaving $\lambda_{\rm tot}$ as a free parameter while $\sigma_{pix}$ and $\mu_{0}$ are constrained to within their uncertainty (Table \ref{tab:fits}), and report the total log-likelihood $\mathcal{LL}= \sum_{i=\textnormal{1}}^{4}\mathcal{LL}_i$.

We find that non-zero values of \sige\ are preferred for DM masses above a few \mevm. 
This is mostly due to the presence of a few pixels with values $>6~\sigma_{pix}$ in the positive tail of the $p$ distribution (Fig. \ref{fig:exthists}), consistent with the higher charge multiplicity expected for larger \mass. However, the presence of a similar tail in the negative side of the $p$ distribution and of similar features in the blank images suggest a noise origin. In Table \ref{tab:pval}, we report the number of pixels found in the negative and positive tails of the $p$ distribution. The thresholds for the tails were chosen appropriately to obtain an expectation of two pixels from the fit with the leakage-only model. There is evidence for an overall excess with comparable numbers on both sides of the distribution and between blank and exposure images.  We conclude that the preference for non-zero values of \sige\ in the fit is due to an imperfect modeling of the extreme tails of the noise distribution. 
Since we do not attempt to parametrize these tails further, more conservative limits are placed when the minimum of the total log-likelihood, $\mathcal{LL_{\rm min}}$, is found at a non-zero value of \sige. For each \mass\ we obtain 90\% C.L. constraints on \sige\ using the test statistic $\Lambda=2(\mathcal{LL}-\mathcal{LL_{\rm min})}$.

\begin{table}[h!]
\centering
\caption{\label{tab:pval} 
Number of pixels in the negative and positive tails of the $p$ distribution, chosen such that there is an expectation of two pixels from the leakage-only fit.}
\begin{ruledtabular}
\begin{tabular}{lcccc}
CCD no.  & 1 & 3 & 4 & 5 \\
& \multicolumn{4}{c}{(negative $p$ tail) / (positive $p$ tail)} \\
\hline
Exposures &  1 / 3 & 2 / 4 & 5 / 5 & 3 / 2
\\
Blanks & 3 / 5 & 4 / 1 & 2 / 1 & 2 / 3
\\
\end{tabular}
\end{ruledtabular}
\end{table}

\begin{figure*}[!ht]
	\includegraphics[width=1.\textwidth]{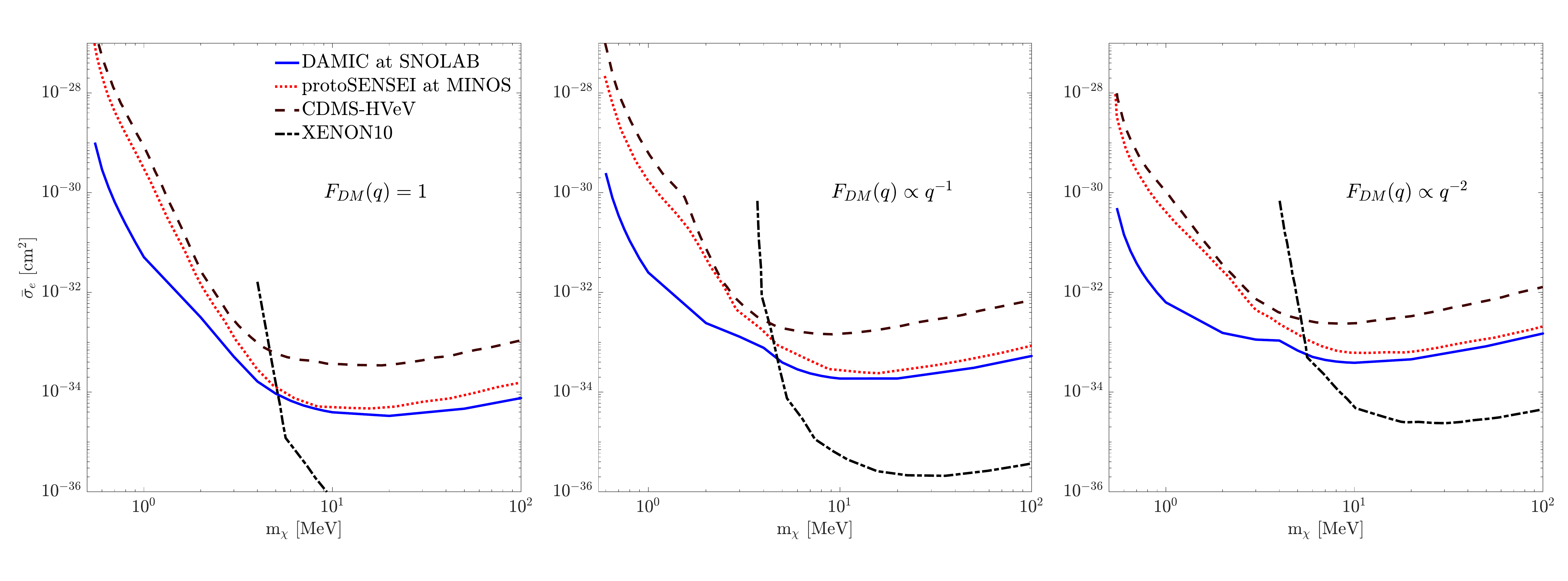}
	\caption{90 \% C.L upper limits on the DM-electron free scattering cross section \sige\ as a function of DM mass \mass\, for $F_{DM} \propto q^{-n}$ ($n=0,1,2$) obtained by DAMIC at SNOLAB (solid line). Current best limits from protoSENSEI at MINOS (dotted line) \cite{abramoff:2019, emken:2019}, CDMS-HVeV surface run (dashed line) \cite{{agnese:2018, *agnese:2019:erratum}}, and an analysis of the XENON10 data (dashed-dotted line) \cite{essig:2017xe} are also shown for comparison.}
	\label{fig:limit}
\end{figure*}

\begin{figure}[!htbp]
\centering
\includegraphics[width=0.48\textwidth]{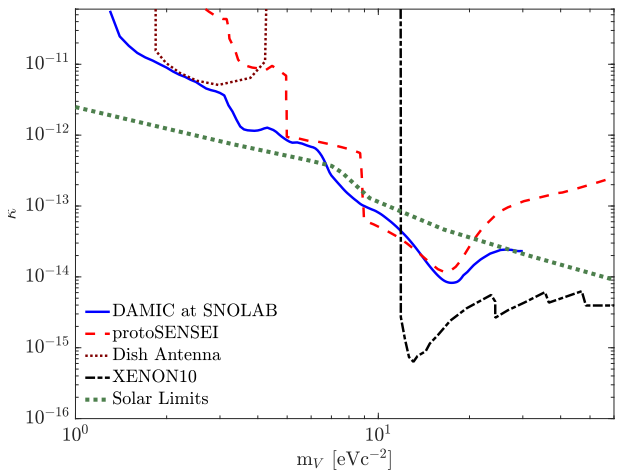}
\caption{90\% C.L. constraints upper limits on the hidden-photon DM kinetic mixing parameter $\kappa$ as a function of the hidden-photon mass $m_V$. Current best direct-detection limits from protoSENSEI at MINOS \cite{abramoff:2019}, an analysis of the XENON10 data \cite{bloch:2017}, a dish antenna \cite{suzuki:2015}, and astrophysical solar limits \cite{bloch:2017, An:2013yfc,*Redondo:2013} are also shown for comparison.}
\label{fig:hplimit}
\end{figure}

The 90\% C.L. constraints on the DM-\electron\ cross section from this analysis are compared in Fig. \ref{fig:limit} to the current best direct-detection limits in Refs. \cite{abramoff:2019, essig:2017xe, agnese:2018,  emken:2019}. Complementary limits for heavier DM masses from noble liquid experiments can be found in Ref. \cite{{aprile:2019:dme, *agnes:2018:dme}}. Note that for a high enough DM-\electron\ cross section the DM flux at SNOLAB would be drastically reduced by interactions in the rock overburden \cite{emken:2019}. However, this region has already been excluded by experiments at shallower sites \cite{emken:2019}. Other constraints from analyses based on astrophysical modifications to the dark matter speed distribution can be found in Ref. \cite{{ema2019light, *an2018directly}}. 

Several checks are performed to evaluate the robustness of the results. A $\pm$5\% systematic uncertainty in the linearity of the calibration constant $\Omega$ changes the limits by $\mp$20\% for \mass\ below few \mevm.  We modify the ionization model by splitting the energy equally between the e-h or assigning it entirely to one of them, with limits changing by $<$10\% for \mass\ below few \mevm. Lastly, we perform the analysis with different central portions of the CCD image, with limits changing by $<$10\%. 

Our previous constraints on hidden-photon dark matter \cite{aguilar:2017} were obtained with a method analogous to the one presented in this Letter. The lower leakage current $\lambda$ and larger exposure of this data set result in more stringent constraints. The corresponding 90\% C.L. upper limits on the hidden-photon kinetic mixing parameter $\kappa$ (also known as $\epsilon$ in literature) as a function of the hidden-photon mass $m_V$ are shown in Fig.~\ref{fig:hplimit}. 

In summary, we have established the best direct-detection limits on dark matter-electron scattering in the mass range of 0.6 \mevm\ to 6 \mevm\ by exploiting the excellent charge resolution and extremely low leakage current of DAMIC CCDs. We also place the best direct-detection constraints on hidden-photon dark matter in the mass range $1.2$\textendash$9$~\evm. Further improvements with the SNOLAB apparatus will be explored by cooling the CCDs to 100~K and improving the light tightness of the cryostat, which may sensibly reduce the leakage current. Improvements of several orders of magnitude are expected with DAMIC-M, a kg-size detector with sub-electron resolution to be installed at the Laboratoire Souterrain de Modane in France \cite{settimo:2018}.  
	
We thank Rouven Essig, Tien-Tien Yu, and Tomer Volansky for assistance with the theoretical background of DM-\electron\ scattering, and Chris Kouvaris for pointing to limit calculations for the overburden. We thank SNOLAB and its staff for support through underground space, logistical and technical services. SNOLAB operations are supported by the Canada Foundation for Innovation and the Province of Ontario Ministry of Research and Innovation, with underground access provided by Vale at the Creighton mine site. We acknowledge the financial support from the following agencies and organizations: Kavli Institute for Cosmological Physics at The University of Chicago through an endowment from the Kavli Foundation; National Science Foundation through Grant No. NSF PHY-1806974; Mexico's Consejo Nacional de Ciencia y Tecnolog\'ia (Grant No. 240666) and Direcci\'on General de Asuntos del Personal Acad\'emico - Universidad Nacional Aut\'onoma 
de M\'exico (Programa de Apoyo a  Proyectos de Investigaci\'on e Innovaci\'on Tecnol\'ogica Grant No. IN108917).

\bibliography{myrefs.bib}

\end{document}